%% file: draft_DtoKSpi0eta.tex
\documentclass[aps,prl,twocolumn,showpacs,amsmath,amssymb, nofootinbib]{revtex4-1}
\newcommand{\BESIIIorcid}[1]{\href{https://orcid.org/#1}{\hspace*{0.1em}\raisebox{-0.45ex}{\includegraphics[width=1em]{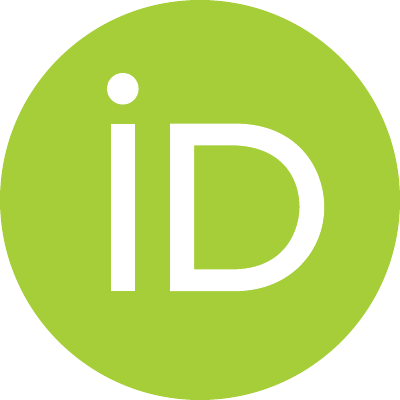}}}}
\usepackage{epsfig}
\usepackage{graphicx}
\usepackage{dcolumn}
\usepackage{bm}
\usepackage{ltablex,booktabs}
\usepackage{overpic}
\usepackage{subfigure}
\usepackage{float}
\usepackage{color}
\usepackage{amsmath}
\usepackage{mathcomp}
\usepackage{mathrsfs}
\usepackage{multirow}
\usepackage{rotating}
\usepackage{amssymb}
\usepackage{gensymb}
\usepackage{amsmath}
\usepackage{tabularx}
\usepackage{tikz}
\usepackage{tikz-feynman}
\usepackage[bookmarksnumbered, pdfstartview=FitH,colorlinks,urlcolor=blue, citecolor=blue,linkcolor=blue] {hyperref}

\begin{document}
\normalsize
\parskip=5pt plus 1pt minus 1pt


\title{\boldmath Study of $\bar{K}^*(892)^0 \eta$ and $K_S^0 a_0(980)^0$ in the $D^{0} \to K_{S}^{0}\pi^0\eta$ decay}

\author{
\begin{small}
  \begin{center}
\input{besauthor}
\end{center}
\end{small}
}
\begin{abstract}
  We perform an amplitude analysis of the decay $D^0 \to K_S^0 \pi^0 \eta$ and
  measure its absolute branching fraction to be
  $(1.016 \pm 0.013_{\text {stat.}} \pm 0.014_{\text {syst.}})\%$. The analysis
  utilizes $20.3~\mathrm{fb}^{-1}$ of $e^{+}e^{-}$ collision data collected at
  a center-of-mass energy of 3.773~GeV with the BESIII detector. The branching
  fraction of the intermediate process $D^0 \rightarrow \bar{K}^*(892)^0 \eta$
  is determined to be
  $(0.73 \pm 0.05_{\text {stat.}} \pm 0.03_{\text {syst.}}) \%$, which is
  $5 \sigma$ smaller than the result measured in $D^0 \to K^- \pi^+ \eta$ by
  the Belle experiment. As a result, the magnitudes of the $W$-exchange and
  QCD-penguin exchange amplitudes are found to be less than half of their
  current estimations.
  Furthermore, we determine $\mathcal{B}(D^0\to K_S^0a_0(980)^0, a_0(980)^0\to \pi^0\eta) = (9.88\pm 0.37_{\rm stat.}\pm 0.42_{\rm syst.})\times10^{-3}$, with  a precision improved by a factor of 4.5
  compared to the world average.
\end{abstract}

\maketitle
Since the first observation of charge-parity~($C\!P$) violation in the charm
sector was announced by the LHCb Collaboration~\cite{LHCb:2019hro}, numerous
theoretical analyses have been conducted to assess whether this observed $C\!P$
violation is consistent with the Standard Model~(SM). Despite the SM's
anticipation of minimal $C\!P$ violation in charm decays, the effects of
non-perturbative quantum chromodynamics~(QCD) caused by final-state
interactions make a consistent and definitive conclusion difficult, and
reliable theoretical models are currently unavailable. Consequently, a
model-independent topological approach~\cite{Chau:1986jb, Cheng:2022vbw} has
been employed to analyze charm decays, deriving the decay amplitudes as a sum
of various topological amplitudes. This method estimates the contribution of
each topological amplitude from  experimental inputs and further  predicts
unmeasured parameters, such as $C\!P$ asymmetries.

\begin{figure}[htbp]
  \centering
  \includegraphics[width=0.235\textwidth]{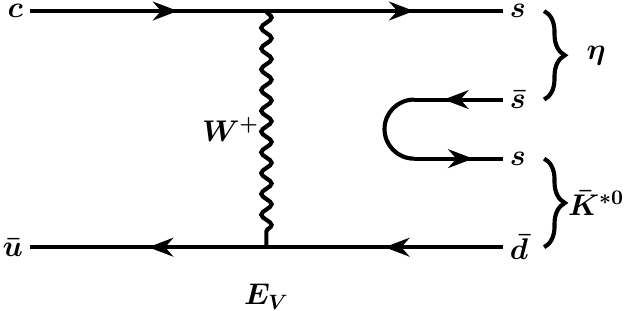}
  \includegraphics[width=0.23\textwidth]{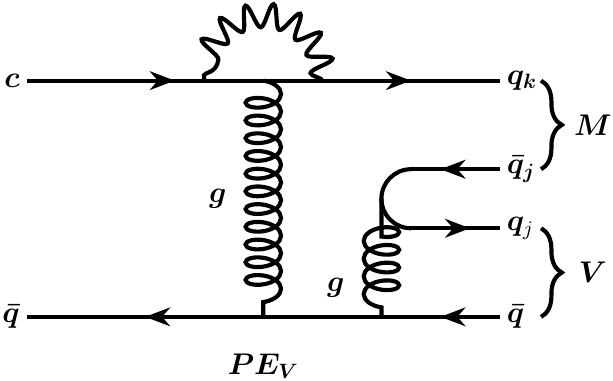}
  \caption{(Left) $W$-exchange diagram for $D^0\to \bar{K}^{*0}\eta$ and
    (Right) QCD-penguin exchange diagram for $D$ decays. $M_1$ shares the same
    spectator quark with the charmed meson, while $M_2$ is an emitted meson.
  }
  \label{feynmen}
\end{figure}

The $C\!P$ violation is more pronounced in singly Cabibbo-suppressed~(SCS)
decays, as their penguin amplitudes provide the weak and strong phase
difference necessary for $C\!P$ violation. To elaborate on the calculation of
SCS decays, the topological approach typically depends on experimental data
from the Cabibbo-favored~(CF) decays to extract various topological amplitudes
and assumes that $W$-exchange amplitudes $E_{V(P)}$ and QCD-penguin exchange
amplitudes $PE_{V(P)}$\footnote{The subscript $V(P)$ for $W$-exchange
amplitudes denotes that the final state antiquark is hadronized to a
vector~(pseudoscalar) meson, while that for QCD-penguin exchange amplitudes
denotes that the spectator quark is hadronized to a vector (pseudoscalar)
meson.}, as depicted in Fig.~\ref{feynmen}, share the same magnitude and
phase~\cite{Cheng:2021yrn}. A notable example is the enhancement of the
predicted $C\!P$ asymmetry in $D^0\to K^+K^-$ from $\sim10^{-4}$ to
$\sim10^{-3}$ due to the inclusion of corresponding $W$-exchange
amplitudes~\cite{Cheng:2021yrn}, which successfully accounts for the LHCb
measurement~\cite{LHCb:2019hro}. Therefore, the accuracy of experimental
studies on charmed meson decays involving $W$-exchange is crucial for the
theoretical estimation of the $C\!P$ asymmetries in the charm sector.

In 2020, the Belle Collaboration measured the branching fraction~(BF) of
the CF $D^0\to \bar{K}^{*0}\eta$ decay to be $(1.41^{+0.13}_{-0.12})\%$ in a
study of $D^0\to K^-\pi^+\eta$~\cite{Belle:2020fbd}, where
$\bar{K}^{*0}\to K^-\pi^+$. This measurement increased the new world average
of the BF, $\mathcal{B}(D^0\to \bar{K}^{*0}\eta)=(1.352\pm0.115)\%$, by 35\%
compared to the previous average~\cite{PDG}. Consequently, the magnitude of
the $W$-exchange amplitude $E_V$\footnote{The subscript $V$ denotes that the
vector meson contains the final-state antiquark $\bar{q}_i$ of the
$W$-exchange amplitude.} is now estimated to be about twice as large compared
to the previous calculation~\cite{Cheng:2021yrn,Cheng:2019ggx}. This
influences the predicted $C\!P$ asymmetries significantly. However, the large
BF from Belle deviates from most theoretical calculations, emphasizing the
urgency of an another measurement to clarify this discrepancy. In this Letter,
we present a BF measurement of $D^0\to \bar{K}^{*0}\eta$ determined from an
amplitude analysis of the neutral decay $D^{0} \to K_S^0\pi^{0}\eta$, where
$\bar{K}^{*0}\to K_S^0\pi^{0}$, using 20.3~fb$^{-1}$ of $e^+e^-$ collision
data collected at a center-of-mass energy
$\sqrt s= 3.773$~GeV~\cite{Ablikim:2013ntc,liangdu, BESIII:2024lbn} with the
BESIII detector \cite{BESIII:detector}. Charge-conjugate states are implied
throughout this Letter.

Furthermore, the process $D^{0} \to K_S^0 a_0(980)^0$ with
$a_0(980)^0\to\pi^{0}\eta$ presents another interesting decay
channel~\cite{Ikeno:2024fjr,Toledo:2020zxj}. The classification of light scalar
particles, such as $a_0(980)$, remains ambiguous despite the success of the
constituent quark model in identifying the nonets of pseudo-scalar, vector, and
tensor mesons. Light scalar particles are considered potential candidates for
compact tetraquarks~\cite{BCKa02,haiyang_4k,BCKa03,BCKa0}, $K\bar K$ bound
states~\cite{Zhang:2022xpf,Feng:2020jvp}, and other possible
states~\cite{Achasov:2017edm,Achasov:2021dvt}, with their production primarily
involving final-state interactions. The substantial final-state interactions
in the hadronic decays of charmed mesons provide an ideal environment for
understanding the structures and properties of light scalar
particles~\cite{Zhang:2024myn,Zhu:2022guw,Zhu:2022wzk,Wang:2021naf,Duan:2020vye}.

\begin{figure}[htbp]
  \centering
  \includegraphics[width=0.157\textwidth]{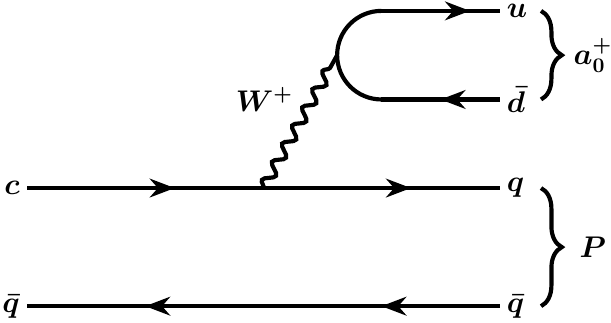}
  \includegraphics[width=0.157\textwidth]{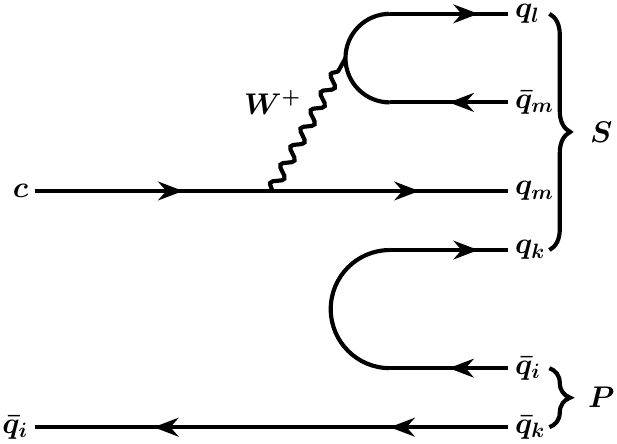}
  \includegraphics[width=0.157\textwidth]{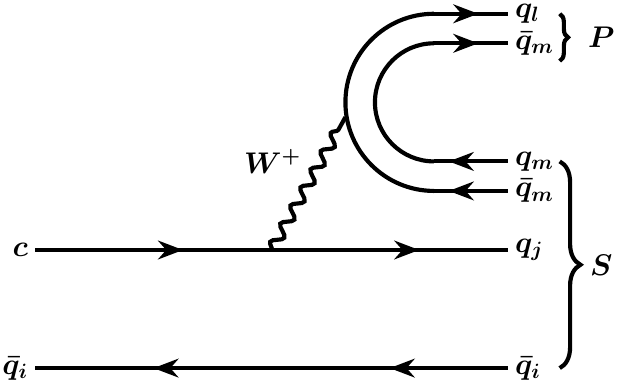}
  \caption{(a) External $W$-emission diagram for $D\to a_0(980)^+ \pi$ in the
    diquark scenario and (b)(c) the two unique $T$-like diagrams specific for
    $D\to SP$ in the tetraquark scenario.}
  \label{fg:Tlike}
\end{figure}

A recent BESIII measurement~\cite{luyu_pipieta} shows that the external
$W$-emission dominates the $D\to a_0(980) \pi$ decays in the diquark scenario,
contrary to expectations of its negligible contribution due to the very small
$a_0(980)$ decay constant. To address this anomaly, Ref.~\cite{haiyang_4k}
introduces two unique $T$-like diagrams specific to the tetraquark scenario,
as shown in Fig.~\ref{fg:Tlike}. These two diagrams lead to a one order of
magnitude difference in the predicted BFs for $D^0\to \bar{K}^0a_0(980)^0$
between the tetraquark and diquark scenarios, making
$D^0\to \bar{K}^0a_0(980)^0$ an excellent process for investigating the nature
of scalar particles. The CLEO Collaboration reported that an accurate amplitude
model could not be established for this decay due to their limited statistics
of 155 events, with a relative uncertainty of approximately $30\%$ in the
BF~\cite{CLEO:2004}. The work reported in this Letter aims to provide a
significantly more precise measurement of this decay. 

A description of the design and performance of the BESIII detector can be found
in Ref.~\cite{BESIII:detector}. Monte Carlo~(MC) simulated data samples are
produced with a {\sc geant4}-based~\cite{geant4} software toolkit, which
includes the geometric description~\cite{detvis} of the BESIII detector and the
detector response. The simulation models the beam energy spread and initial
state radiation~(ISR) in the $e^+e^-$ annihilations with the generator
{\sc kkmc}~\cite{ref:kkmc}. The inclusive MC sample includes the production of
$D\bar{D}$ pairs, the non-$D\bar{D}$ decays of the $\psi(3770)$, the ISR
production of the $J/\psi$ and $\psi(3686)$ states, and the continuum processes
incorporated in {\sc kkmc}. All particle decays are modeled with
{\sc evtgen}~\cite{ref:evtgen} using BFs either taken from the Particle Data
Group~(PDG)~\cite{PDG:2024}, when available, or otherwise estimated with
{\sc lundcharm}~\cite{ref:lundcharm}. Final-state radiation from charged
final-state particles is incorporated using {\sc photos}~\cite{photos}.

A tag technique is used to measure the absolute BF and to provide samples with
high purity for amplitude analyses. In the single-tag~(ST) sample, only the
$\bar{D}^{0}$ meson is reconstructed through one of the three decay
modes~(tag modes): $K^+ \pi^-$, $ K^+ \pi^- \pi^0$, and
$K^+ \pi^- \pi^+ \pi^-$, without any requirement on the remaining measured
tracks and electromagnetic calorimeter
showers~\cite{MA_BF_etaX, BESIII:2023htx,Ke:2023qzc}. In the double-tag~(DT)
sample, the $D^{0}$ meson is reconstructed through the signal decay mode
$K_S^0\pi^0\eta$ ,and the associated $\bar{D}^{0}$ through one of the tag
modes. The selection criteria for the final-state particles are the same as
those in Refs.~\cite{MA_BF_etaX, BESIII:2023htx}.

The $D$ mesons are selected using the energy difference
$\Delta E = E_{D} - E_{\rm beam}$ and the beam-constrained mass
$M_{\rm BC} = \sqrt{E^{2}_{\rm beam}-|\vec{p}_{D}|^{2}}$, where
$E_{\rm beam}$ is the beam energy and $\vec{p}_{D}$ and $E_{D}$ are the momentum
and the energy of the $D$ candidate in the $e^+e^-$ rest frame, respectively.
In case of multiple candidates, the one with the minimum $|\Delta{E}|$ is
chosen.

The DT sample is used in the amplitude analysis, with the requirements that $D$
candidates satisfy $1.859<M_{\rm BC}<1.873$~GeV/$c^2$ and the $\Delta{E}$
window listed in Table~\ref{tab:ST_DT_efficiency} for the tag modes, and
$1.858<M_{\rm BC}<1.871$~GeV/$c^2$ and $-50<\Delta{E}<45$~MeV for the signal
mode. Since both $\pi^0$ and $\eta$ are reconstructed through their di-photon
decays, the four photons from decay modes with two $\pi^0$ mesons could be
incorrectly paired even if the photons are correctly reconstructed. Events are
therefore  rejected if any combination of both pairs of photons falls within
the $\pi^0$ mass region: $0.115<M_{\gamma\gamma}<0.150$~GeV/$c^2$. A total of
6078 DT events are selected for the amplitude analysis with a signal purity
$\omega_{\rm sig}$ of $(94.6\pm 0.5)\%$. The purity is determined from a
two-dimensional~(2D) unbinned maximum-likelihood fit to the distribution of
$M_{\rm BC}$ of the tag $\bar{D}^0$ versus that of the signal $D^0$. The 2D fit
method is consistent with the one described in Ref.~\cite{MA_BF_etaX}.

\begin{figure}[htbp]
  \centering
  \includegraphics[width=0.235\textwidth]{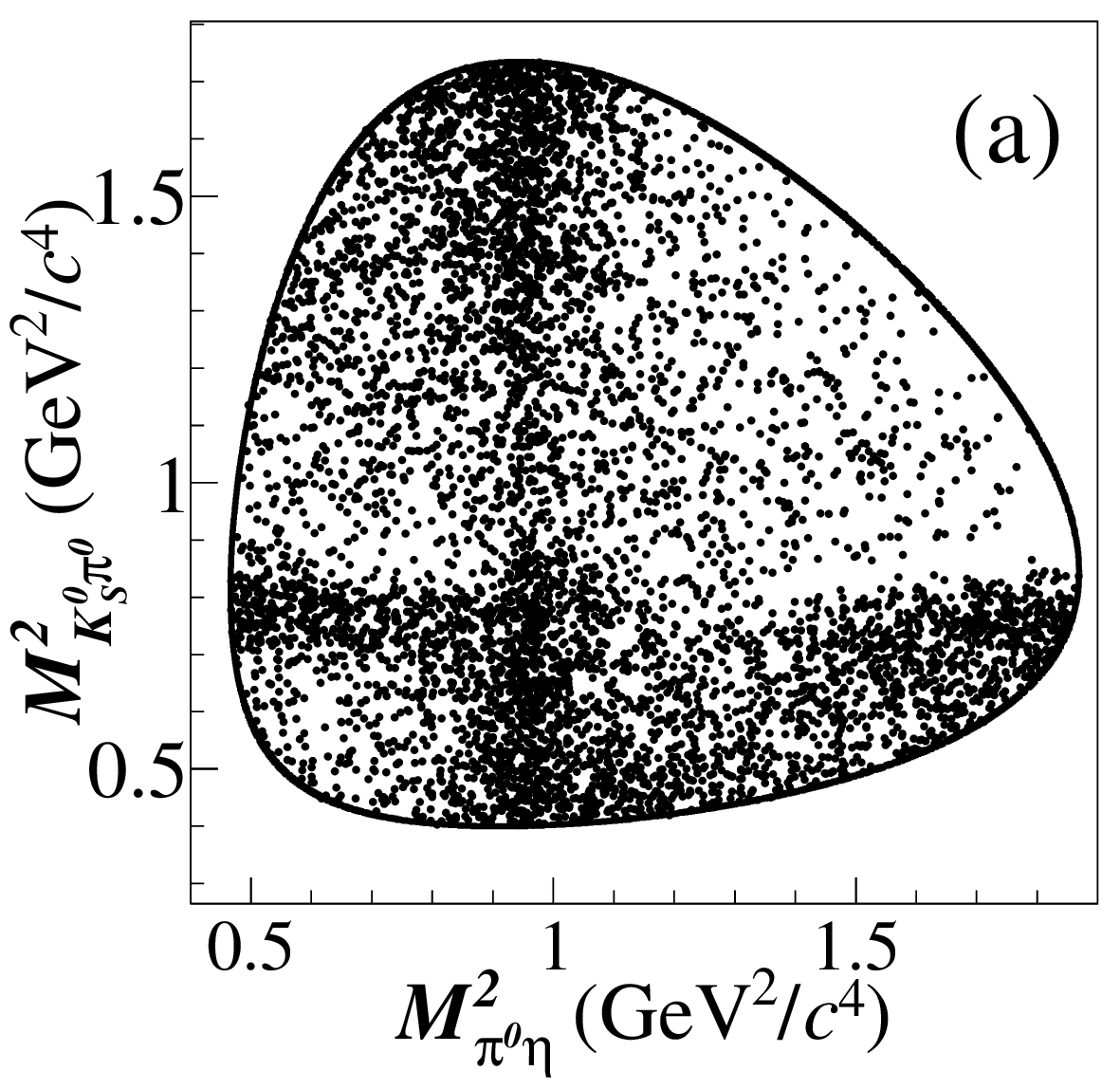}
  \includegraphics[width=0.235\textwidth]{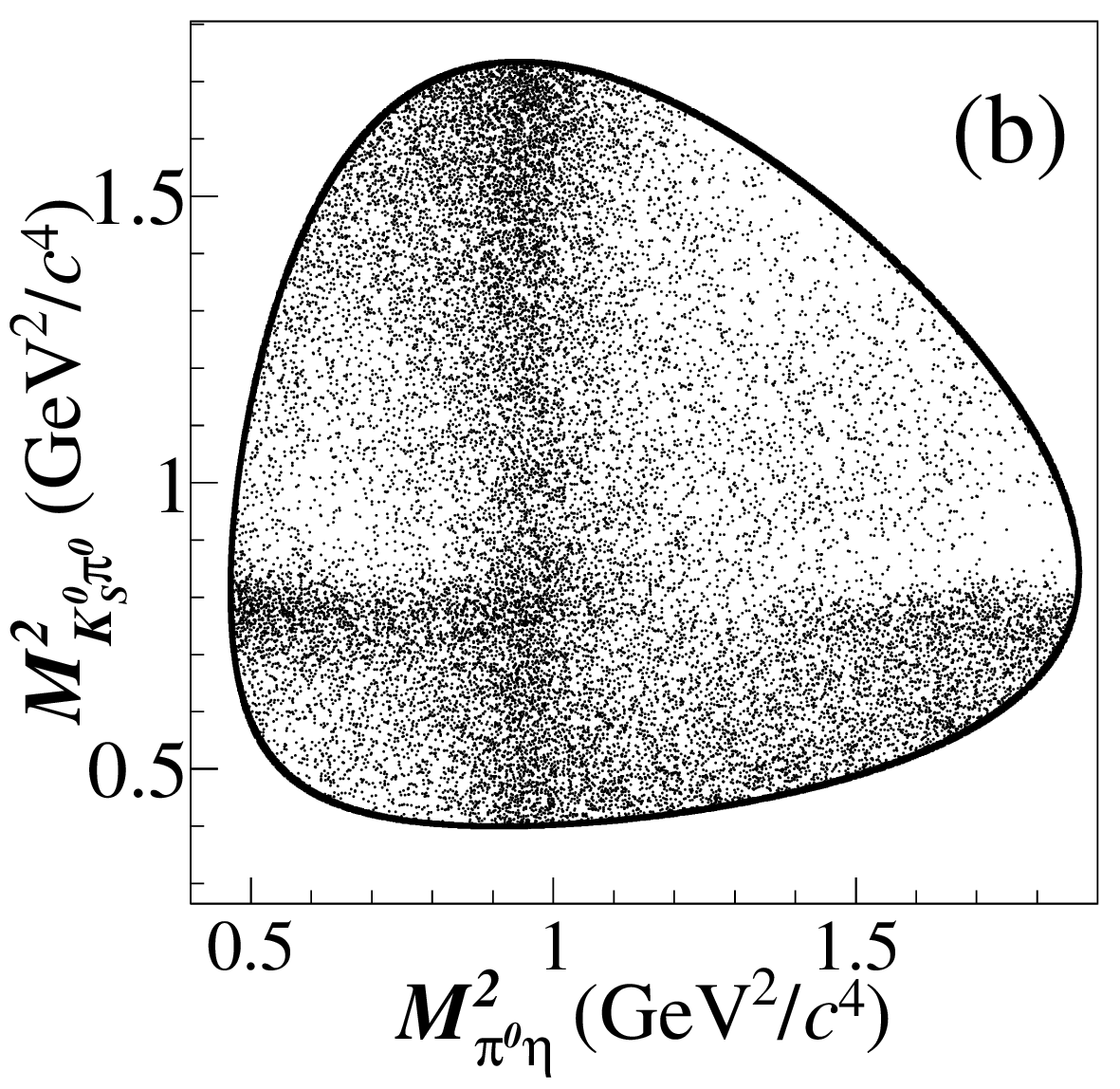}
  \caption{Dalitz plots of $M^{2}_{\pi^0\eta}$ versus $M^{2}_{K_S^0\eta}$
    of the $D^0\to K^0_S\pi^0\eta$ candidates for (a) the data sample and (b)
    the signal MC sample generated according to the amplitude analysis
    results.}
  \label{dalitz}
\end{figure}

Since the amplitude analysis requires a sample with good momentum resolution
and all candidates falling within the phase-space boundary, a  four-constraint
kinematic fit is performed, constraining the invariant masses of
$(\gamma\gamma)_{\eta}$, $(\gamma\gamma)_{\pi^{0}}$,
$(\pi^{+}\pi^{-})_{K_S^0}$ and $D^{0}$ candidates to their known
masses~\cite{PDG:2024}. The Dalitz plot of the $D^0\to K^0_S\pi^0\eta$
candidates for the data sample is shown in Fig.~\ref{dalitz}(a).
Figure~\ref{dalitz}(b) shows the Dalitz plot for the signal MC sample generated
based on the results of the amplitude analysis.

The composition of intermediate processes is determined by an unbinned 
maximum-likelihood fit with the likelihood function $\mathcal{L}$ constructed
by a signal probability density function~(PDF):
$\mathcal{L} = \prod_{k}^{N_{\rm D}}f_{S}(p_{j}^{k})$. The signal PDF $f_{S}$
depends on the momenta $p$ of the final-state particles, $j$ runs over the
three final-state particles, $k$ runs over data events, and $N_{\rm D}$ is the
number of candidate events in data. 
The signal PDF is written as
\begin{eqnarray}\begin{aligned}
  f_{S}(p_{j}) = \frac{\epsilon(p_{j})\left|\mathcal{M}(p_{j})\right|^{2}R_{3}}{\int \epsilon(p_{j})\left|\mathcal{M}(p_{j})\right|^{2}R_{3}\,dp_{j}}\,, \label{signal-PDF}
\end{aligned}\end{eqnarray}
where $\epsilon(p_{j})$ is the detection efficiency and $R_{3}$ is the element
of three-body phase space. The total amplitude $\mathcal{M}\left(p_j\right)$ is
the coherent sum of the individual amplitudes of intermediate processes,
expressed as $\sum_n \rho_n e^{i \phi_n} \mathcal{A}_n$, where $\rho_n$ and
$\phi_n$ denote the magnitude and phase of the amplitude $\mathcal{A}_n$ for
the $n$-th intermediate process, respectively. The amplitude $\mathcal{A}_{n}$
is the product of the spin factor~\cite{covariant-tensors}, the Blatt-Weisskopf
barriers of the intermediate resonance and the $D^{0}$
meson~\cite{PhysRevD.83.052001}, and  the propagators for the intermediate
resonances. The propagator of $a_0(980)^0$ is parameterized using  a Flatt\'e
formula with the parameters given in Ref.~\cite{a0980para}. The propagators of
$a_2(1320)^0$, $\bar{K}_0^*(892)^0$, $\bar{K}_0^*(1410)^0$, $\bar K^{*}(1680)^0$
and $\bar K_2^{*}(1980)^0$ are parameterized as a relativistic
Breit-Wigner~(RBW) function~\cite{RBW} using parameters obtained in
Ref.~\cite{K1430para}. The $K_S^0\pi^0$ $S$-wave is modeled by the LASS
parameterization and the relevant parameters are fixed to the BaBar and Belle
results~\cite{KpiLASS, Belle:2020fbd}. The normalization integral term in the
denominator is calculated using an MC integration~\cite{BESIII:2022kbq}. A
background PDF
\begin{equation}
  f_B(p_j)=\frac{B(p_j)R_3(p_j)}{\int{B(p_j)R_3(p_j)}\mathrm{d}p_j},
  \label{eq:bkglikelihood}
\end{equation}
is incoherently added to the signal PDF, where the background amplitude $B$ is
simulated with the inclusive MC samples, and the likelihood function is
modified as~\cite{Langenbruch:2019nwe}:
\begin{equation}
  \ln{\mathcal{L}} = \begin{matrix}\sum\limits_{k}^{N_{\rm D}} \ln [\omega_{\rm sig}f_{S}(p^k_{j})+(1-\omega_{\rm sig})f_{B}(p^k_j)]\end{matrix}.
  \label{eq:likefinal}
\end{equation}

During the fit, the magnitude and phase of the amplitude
$D^{0} \to K_S^0a_0(980)^{0}$ are fixed to be 1.0 and 0.0, respectively, while
those of the other amplitudes are floating. Various combinations of amplitudes
for intermediate resonances are tested. The nominal combination includes
$D^{0}\to K_S^0a_0(980)^{0}$, $D^0\to K_S^0a_2(1320)^0$,
$D^{0}\to\bar{K}^{*}(982)^{0}\eta$, $D^{0}\to\bar{K}^{*}(1410)^{0}\eta$,
$D^0\to (K_S^0\pi^0)_{S\rm -wave}\eta$, $D^0\to \bar K^{*}(1680)^0\pi^0$, and
$D^0\to \bar K_2^{*}(1980)^0\pi^0$, for which all amplitudes have a
significance greater than 5$\sigma$. Other possible intermediate processes,
including $D^0\to\bar{K}^{*}_0(1430)^0\eta$, $D^0\to\bar{K}^{*}_2(1430)^0\eta$,
$D^0\to\bar{K}^{*}(1680)^0\eta$, $D^0\to\bar{K}^{*}_2(1980)^0\eta$,
$D^0\to\bar{K}^{*}_0(1430)^0\pi^0$, $D^0\to\bar{K}^{*}_2(1430)^0\pi^0$, and
$D^{0} \to (K_S^0\eta)_{S-{\rm wave}}\pi^0$, have been tested, but have no
significant contributions.

The relative contribution of the $n$-th intermediate process is quantified by a
fit fraction~(FF) defined as
\begin{eqnarray}\begin{aligned}
  {\rm FF}_{n} = \frac{\int \left|\rho_{n}e^{i\phi_{n}}\mathcal{A}_{n}\right|^{2}R_3(p_j)\mathrm{d}p_j}{\int\left|\mathcal{M}\right|^{2}R_3(p_j)\mathrm{d}p_j}\,. \label{Fit-Fraction-Definition}
\end{aligned}\end{eqnarray}
The statistical uncertainties are determined by randomly sampling distributions
of FFs according to the covariance matrix. 

\begin{figure*}[!htbp]
  \centering
  \includegraphics[width=0.3\textwidth]{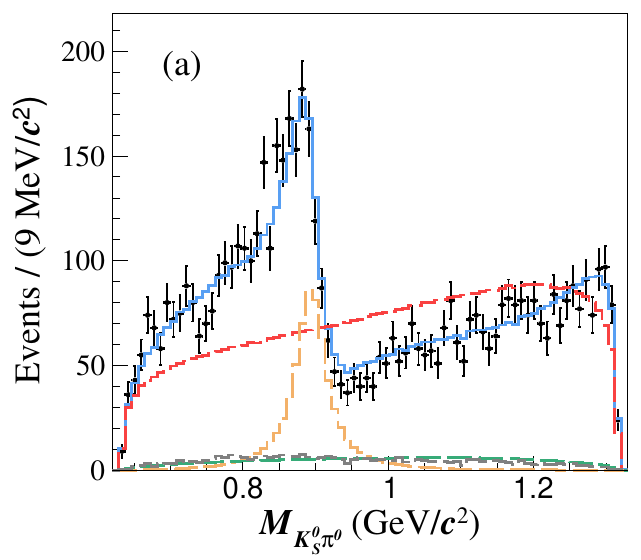}
  \includegraphics[width=0.3\textwidth]{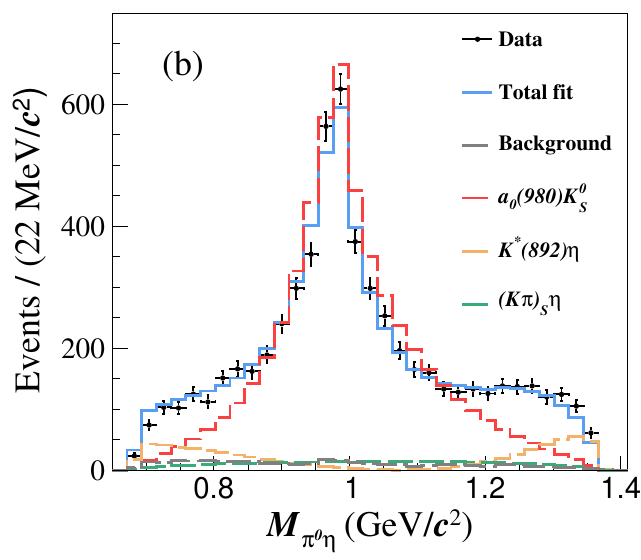}
  \includegraphics[width=0.3\textwidth]{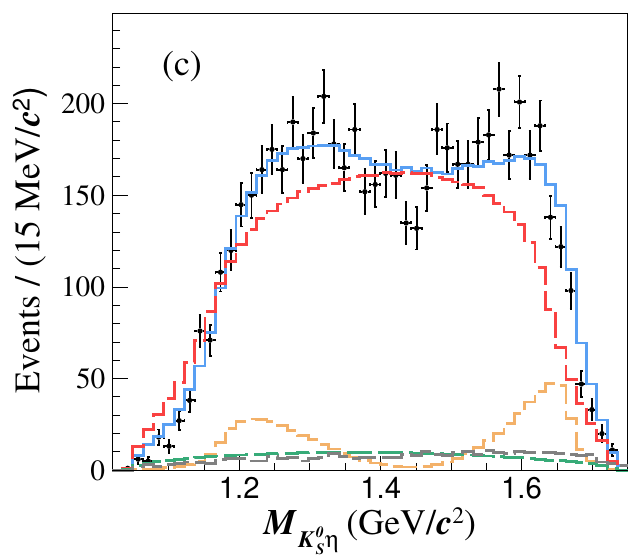}
  \caption{Projections of the invariant masses of the (a) $K_S^0\pi^0$, (b)
    $\pi^0\eta$ and (c) $K_S^0\eta$ systems from the nominal fit. The data are
    represented by black dots with error bars, the fit results by solid blue
    curves, the backgrounds by dashed grey curves, and the intermediate
    processes by the other colored curves. 
  }
  \label{dalitz-projection}
\end{figure*}

Figure~\ref{dalitz-projection} shows the mass projections of the fit result.
Table~\ref{fit-result} lists the fitted phases, FFs, and statistical
significances for the intermediate processes. The systematic uncertainties of
the amplitude analysis are estimated as described below.

\begin{table*}[htbp]
  \caption{Phases, FFs, statistical significances, and BFs for the significant
    intermediate processes. The BFs are calculated as
    $\mathcal{B}={\rm FF}\times \mathcal{B}(D^0 \to K_S^0\pi^0\eta)$. The first
    uncertainties are statistical, while the second are systematic. The total
    of the FFs is 122.6\%, exceeding 100\% due to interference effects.}
  \label{fit-result}
  \begin{center}
    \begin{tabular}{lccc}
      \hline
      \hline
      Amplitude & Phase $\phi$ (rad) & FF~(\%)  & BF~$(10^{-3})$\\
      \hline
      $D^0\to K_S^0a_0(980)^0, a_0(980)^0\to \pi^0\eta$           
      & 0.0(fixed)                      
      & $93.2\pm3.3\pm3.2$   
      & $9.88\pm 0.37\pm 0.42$\\

     $D^0\to K_S^0a_2(1320)^0, a_2(1320)^0\to \pi^0\eta$
      & $-5.41\pm0.24\pm0.22$
      & $0.8\pm0.1\pm0.1$ 
      & $0.08 \pm 0.11 \pm 0.11$\\
      
      $D^0\to \bar K^{*}(892)^0\eta, \bar K^{*}(892)^0\to K^0_S\pi^0$
      & $-2.70\pm0.09\pm0.06$
      & $11.5\pm0.8\pm1.0$ 
      & $1.22 \pm 0.09 \pm 0.11$\\

      $D^0\to \bar K^{*}(1410)^0\eta, \bar K^{*}(1410)^0\to K^0_S\pi^0$
      & $-1.66\pm0.07\pm0.08$  
      & $5.3\pm1.3\pm1.5$
      & $0.56 \pm 0.14 \pm 0.16$\\
      
      $D^0\to (K_S^0\pi^0)_{S\rm-wave}\eta$
      & $-4.05 \pm0.10 \pm0.04$     
      & $6.0 \pm0.9\pm1.0$   
      & $0.64 \pm 0.10 \pm 0.10$\\

      $D^0\to \bar K^{*}(1680)^0\pi^0, \bar K^{*}(1680)^0\to K^0_S\eta$
      & $-3.39\pm0.13\pm0.09$  
      & $1.8\pm0.7\pm0.5$   
      & $0.19 \pm 0.07 \pm 0.05$\\

       $D^0\to \bar K_2^{*}(1980)^0\pi^0, \bar K_2^{*}(1980)^0\to K^0_S\eta$
      & $-4.59\pm0.08\pm0.04$ 
      & $4.0\pm1.4\pm1.5$   
      & $0.45\pm 0.15 \pm 0.16$\\
      \hline
      \hline
    \end{tabular}
  \end{center}
\end{table*}
The systematic uncertainty due to the $a_0(980)^0$ lineshape is estimated by 
varying the mass and coupling constants in the Flatt\'e propagator by
$\pm 1\sigma$ and utilizing a form of three-channel-coupled Flatt\'e formula
with the inclusion of the extra channel $\pi\eta^\prime$ according to
Ref.~\cite{a0980para}. The $a_2(1320)^0$, $\bar{K}_0^*(892)^0$,
$\bar{K}_0^*(1410)^0$, $\bar K^{*}(1680)^0$ and $\bar K_2^{*}(1980)^0$
lineshapes are estimated by varying the masses and widths in the RBW by
$\pm 1\sigma$ according to Ref.~\cite{K1430para}. The
$(K_S^0\pi^0)_{S\rm-wave}$ lineshapes are determined by varying the parameters
of the LASS parameterization~\cite{KpiLASS} by $\pm 1\sigma$. The effective
radius of the Blatt-Weisskopf barrier~\cite{PhysRevD.83.052001} is determined by
varying the effective radius within the range $[2.0, 4.0]$~GeV for intermediate
resonances and $[4.0, 6.0]$~GeV for $D^0$ mesons. 

To determine the systematic uncertainty related to the background estimation,
we vary the signal purity, \textit{i.e.} $\omega_{\rm sig}$ in
Eq.~(\ref{eq:likefinal}), by $\pm 1\sigma$. The uncertainty associated with the
detector acceptance difference between MC samples and data is determined by
reweighting $\epsilon(p_{j})$  in Eq.~(\ref{signal-PDF}) with different
reconstruction efficiencies of $K_S^0$, $\pi^0$, and $\eta$ according to their
uncertainties. The changes of the fit results are taken as the systematic
uncertainties. 

For the BF measurement, we use MC samples generated using the results of the
amplitude analysis to determine the overall efficiency.  We moderately widen
the $\Delta{E}$ window from $-45<\Delta{E}<50$~MeV to $-57<\Delta{E}<45$~MeV
and $1.8365<M_{\rm BC}<1.8865$~GeV/$c^2$ for the signal mode. The remaining
selection criteria are the same as those used in the amplitude analysis, except
for the absence of the $M_{\mathrm{BC}}$ requirements and the five-constraint
kinematic fit. The absolute BF of $D^0 \to K_S^0\pi^0\eta$ is determined by 
\begin{equation}
  \mathcal{B}_{\rm sig}=\frac{N^{\rm DT}}{\mathcal{B}_{\rm sub} \sum_{\alpha}{N^{\rm ST}_{\alpha}}\epsilon^{\rm DT}_{\alpha,\rm sig}/\epsilon^{\rm ST}_{\alpha}[1-{\frac{2r_{\alpha}R_{\alpha}{\rm cos}{\delta_{\alpha}}}{1+r_{\alpha}^{2}}}(2F_{+}-1)]},
  \label{abs:bf_qcc}
\end{equation}
where $\alpha$ denotes the tag mode. The ST yields $N_{\alpha}^{\rm ST}$ and
ST efficiencies $\epsilon_{\alpha}^{\rm ST}$ are obtained by fitting the
$M_{\rm BC}$ distributions of the ST candidates in data and inclusive MC
samples, respectively.  Here, the signal shape is modeled with the
MC-simulated shape convolved with a double-Gaussian function and the background
is parameterized as the ARGUS function~\cite{argus} with an endpoint parameter
fixed at 1.8865~GeV/$c^{2}$. The DT yield $N^{\rm DT}$ is found to be
$6633\pm 89$ from a fit to the distribution of $M_{\rm BC}$ of the tag
$\bar{D}^0$ versus  the signal $D^0$. We utilize the same fitting
method~\cite{MA_BF_etaX} as in the amplitude analysis to obtain the DT yield.

To account for the influence of peaking background $D^0\to K_S^0\pi^0\pi^0$, we
use MC simulations to estimate its impact. The corresponding DT efficiencies
$\epsilon_{\alpha}^{\rm DT}$ are obtained by analyzing the signal MC samples,
with the signal events for $D^0\to K_S^0\pi^0\eta$ generated based on the
results of the amplitude analysis. All the numerical results are summarized in
Table~\ref{tab:ST_DT_efficiency}. The parameters $r_{\alpha}$, $R_{\alpha}$,
$\delta_{\alpha}$ and $F_+$ are introduced to take into account
quantum-correlation effects. Their values are taken from
Refs.~\cite{tag0,tag13}. The BF of $D^0 \to K_S^0\pi^0\eta$ is determined to be
$(1.016\pm0.013_{\rm stat.}\pm0.014_{\rm syst.})\%$. 

\begin{table*}
  \caption{ST yields in data~($N^\textrm{ST}_{\alpha}$), ST
    efficiencies~($\epsilon^\textrm{ST}_{\alpha}$) and DT
    efficiencies~($\epsilon^\textrm{DT}_{\alpha}$) for tag mode $\alpha$.
    The uncertainty of $N^\textrm{ST}_{\alpha}$ is statistical, those of
    $\epsilon^\textrm{ST}_{\alpha}$ and $\epsilon^\textrm{DT}_{\alpha}$ are
    negligible. The efficiencies do not include the BFs of
    final-state particle decays.}
  \label{tab:ST_DT_efficiency}
  \begin{tabular} {l |c c c c}  
    \hline \hline 
    Tag mode &$\Delta E$~(MeV) & $N_\textrm{tag}$~($\times10^3$)  & $\epsilon_\textrm{tag}~(\%)$ & $\epsilon^\textrm{DT}_{\alpha}~(\%)$\\ 
    \hline    
    $ K^+ \pi^-$             &$(-27, 27)$ & $38208 \pm 20$           & $66.67\pm 0.01$             & $10.63\pm0.01$ \\
    $ K^+ \pi^- \pi^0$       &$(-62, 49)$ & $79265 \pm 32$           & $37.92\pm0.01$             & $5.30\pm 0.01$ \\
    $ K^+ \pi^- \pi^+ \pi^-$ &$(-26, 24)$ & $51401 \pm 26$           &$42.20\pm 0.01$             & $5.75\pm 0.01$ \\
    \hline \hline 
  \end{tabular}
\end{table*}

The systematic uncertainties arise from the sources discussed below and are
estimated relative to the measured BFs. The uncertainties in the total ST
yields come from the $M_{\rm BC}^{\rm tag}$ fits to the ST $\bar D^{0}$
candidates, which were determined as 0.3$\%$ by varying the fitting signal
and background shapes. The systematic uncertainties associated with the
$K^0_S$ reconstruction are found to be 0.2$\%$ per $K^0_S$ by studying
the DT samples of $D^0 \to K_S^0\pi^+\pi^-$, $D^0 \to K_S^0\pi^+\pi^-\pi^0$,
and $D^0 \to K_S^0\pi^0$. The systematic uncertainties caused by the
$\pi^0/ \eta$ reconstruction are assigned as 0.3\% and 0.5\% by studying the
DT samples of $D^0 \to K^-\pi^+$ versus $D^0 \to K^-\pi^+\pi^0$. The
uncertainties due to the quoted BFs of $K^0_S \to \pi^+\pi^-$,
$\eta \to \gamma \gamma$, and $\pi^0 \to \gamma \gamma$ decays are 0.07$\%$,
0.18$\%$, and 0.34$\%$, respectively. To estimate the systematic uncertainty in
the 2D fit, we repeat the fits by varying the signal shape,  and the endpoint
of the ARGUS function by $\pm 0.2$~MeV, the fixed peaking background yield of
$D^0\to K_S^0\pi^0\pi^0$~($\pm1\sigma$ of the quoted BF). The systematic
uncertainty due to the $\Delta E$ requirement is assigned to be 0.3$\%$, which
is the largest efficiency difference with and without smearing the data-MC
resolution of $\Delta E$ for signal MC events. The uncertainty from the signal
MC model based on the results of the amplitude analysis is studied by varying
the fit parameters according to the covariance matrix. The change of signal
efficiency is estimated to be 0.3\%. The uncertainties from the
quantum-correlation parameters $r_{\alpha}$, $R_{\alpha}$, $\delta_{\alpha}$
and $F_+$ are propagated according to results of Refs.~\cite{tag0,tag13},
giving a value of 0.5\%. 

In summary, we have performed an amplitude analysis and BF measurement of
$D^0 \to K_S^0\pi^0\eta$ with a much improved precision using 20.3~fb$^{-1}$ of
$e^+e^-$ collision data taken at $\sqrt s= 3.773$~GeV with the BESIII detector.
We determine
$\mathcal{B}(D^0 \to K_S^0\pi^0\eta) = (1.016\pm0.013_{\rm stat.}\pm0.014_{\rm syst.})\%$.

The dominant intermediate resonances are $a_0(980)^0$ and $\bar{K}^*(892)^0$.
The BFs for the intermediate processes are summarized in
Table~\ref{fit-result}. The BF of $D^0 \to\bar{K}^*(892)^0 \eta$ is determined
to be $(0.73 \pm 0.05_{\rm stat.} \pm 0.03_{\rm syst.}) \%$, which is
$5 \sigma$ lower than the measurement by Belle~\cite{Belle:2020fbd},
$(1.41_{-0.12}^{+0.13} )\%$. Our result is consistent with various theoretical
predictions of
$(0.51-0.93)\%$~\cite{qinqin_FSI,qinqin_pole, Cheng:2010ry,Cheng:2019ry}.
According to our findings, the magnitudes of the $W$-exchange amplitudes $E_V$
and QCD-penguin exchange amplitudes $P E_V$ should be less than half of their
currently estimated values~\cite{Cheng:2021yrn}. This substantial revision, as
the major source, will significantly impact the predicted SM $C\!P$ violation
for $D\to VP$ modes~\cite{Cheng:2021yrn}. The nature of this impact can be
anticipated by comparing the predictions in Table VI of
Ref.~\cite{Cheng:2021yrn} and Table IX of Ref.~\cite{Cheng:2019ry}. For
instance, the predicted $C\!P$ asymmetry in decays such as $D \to \bar{K}^0K^*$
is decrease markedly, potentially shifting from the 1.07 level to a
significantly lower value, $ -0.34$~\cite{Cheng:2021yrn,Cheng:2019ry}.

Moreover, we measure the BF
$\mathcal{B}(D^0 \to K_S^0 a_0(980)^0 ,$ $a_0(980)^0 \to \pi^0 \eta)$ $= \left(9.88 \pm 0.37_{\text {stat.}} \pm 0.42_{\text {syst. }}\right) \times 10^{-3}$. This BF is consistent with the
prediction in Ref.~\cite{haiyang_4k} assuming a tetraquark scenario,
$(9.5-15)\times 10^{-3}$, but disfavors that in the diquark scenario,
$6.5\times10^{-4}$.


\acknowledgments
The BESIII Collaboration thanks the staff of BEPCII and the IHEP computing center for their strong support. This work is supported in part by National Key R\&D Program of China under Contracts Nos. 2023YFA1606000, 2023YFA1606704; National Natural Science Foundation of China (NSFC) under Contracts Nos. 11635010, 11735014, 11935015, 11935016, 11875054, 11935018, 12205384, 12025502, 12035009, 12035013, 12061131003, 12192260, 12192261, 12192262, 12192263, 12192264, 12192265, 12221005, 12225509, 12235017, 12361141819; the Chinese Academy of Sciences (CAS) Large-Scale Scientific Facility Program; the CAS Center for Excellence in Particle Physics (CCEPP); Joint Large-Scale Scientific Facility Funds of the NSFC and CAS under Contract Nos. U2032104, U1832207; The Excellent Youth Foundation of Henan Scientific Commitee under Contract No. 242300421044; 100 Talents Program of CAS; The Institute of Nuclear and Particle Physics (INPAC) and Shanghai Key Laboratory for Particle Physics and Cosmology; German Research Foundation DFG under Contract No. FOR5327; Istituto Nazionale di Fisica Nucleare, Italy; Knut and Alice Wallenberg Foundation under Contracts Nos. 2021.0174, 2021.0299; Ministry of Development of Turkey under Contract No. DPT2006K-120470; National Research Foundation of Korea under Contract No. NRF-2022R1A2C1092335; National Science and Technology fund of Mongolia; National Science Research and Innovation Fund (NSRF) via the Program Management Unit for Human Resources \& Institutional Development, Research and Innovation of Thailand under Contracts Nos. B16F640076, B50G670107; Polish National Science Centre under Contract No. 2019/35/O/ST2/02907; Swedish Research Council under Contract No. 2019.04595; The Swedish Foundation for International Cooperation in Research and Higher Education under Contract No. CH2018-7756; U. S. Department of Energy under Contract No. DE-FG02-05ER41374.

\end{document}

%% file: besauthor.tex
M.~Ablikim$^{1}$\BESIIIorcid{0000-0002-3935-619X},
M.~N.~Achasov$^{4,b}$\BESIIIorcid{0000-0002-9400-8622},
P.~Adlarson$^{76}$\BESIIIorcid{0000-0001-6280-3851},
X.~C.~Ai$^{81}$\BESIIIorcid{0000-0003-3856-2415},
R.~Aliberti$^{35}$\BESIIIorcid{0000-0003-3500-4012},
A.~Amoroso$^{75A,75C}$\BESIIIorcid{0000-0002-3095-8610},
Q.~An$^{72,58,\dagger}$,
Y.~Bai$^{57}$\BESIIIorcid{0000-0001-6593-5665},
O.~Bakina$^{36}$\BESIIIorcid{0009-0005-0719-7461},
Y.~Ban$^{46,g}$\BESIIIorcid{0000-0002-1912-0374}
% ...
\\
\vspace{0.2cm}
(BESIII Collaboration)\\
\vspace{0.2cm} {\it
$^{1}$ Institute of High Energy Physics, Beijing 100049, People's Republic of China\\
% ...
$^{4}$ Budker Institute of Nuclear Physics SB RAS (BINP), Novosibirsk 630090, Russia\\
% ...
$^{35}$ Johannes Gutenberg University of Mainz, Johann-Joachim-Becher-Weg 45, D-55099 Mainz, Germany\\
$^{36}$ Joint Institute for Nuclear Research, 141980 Dubna, Moscow region, Russia\\
% ...
$^{46}$ Peking University, Beijing 100871, People's Republic of China\\
% ...
$^{57}$ Southeast University, Nanjing 211100, People's Republic of China\\
$^{58}$ State Key Laboratory of Particle Detection and Electronics, Beijing 100049, Hefei 230026, People's Republic of China\\
% ...
$^{72}$ University of Science and Technology of China, Hefei 230026, People's Republic of China\\
% ...
$^{75}$ University of Turin and INFN, (A)University of Turin, I-10125, Turin, Italy; (B)University of Eastern Piedmont, I-15121, Alessandria, Italy; (C)INFN, I-10125, Turin, Italy\\
$^{76}$ Uppsala University, Box 516, SE-75120 Uppsala, Sweden\\
% ...
$^{81}$ Zhengzhou University, Zhengzhou 450001, People's Republic of China\\

\vspace{0.2cm}
$^{\dagger}$ Deceased\\
% ...
$^{b}$ Also at the Novosibirsk State University, Novosibirsk, 630090, Russia\\
% ...
$^{g}$ Also at State Key Laboratory of Nuclear Physics and Technology, Peking University, Beijing 100871, People's Republic of China\\
% ...

}